\documentclass[%
superscriptaddress,
twocolumn,
 amsmath,amssymb,
 aps,
 pra,
]{revtex4-1}

\usepackage{graphicx}
\usepackage{dcolumn}
\usepackage{bm}
\usepackage{xcolor}
\usepackage{hyperref}

\usepackage{graphicx}
\usepackage{amsmath}
\usepackage{bbold}
\usepackage[T1]{fontenc}

\newcommand{\beq}{\begin{equation}}
\newcommand{\eeq}{\end{equation}}
\newcommand{\beqs}{\begin{equation*}}
\newcommand{\eeqs}{\end{equation*}}
\newcommand{\beqa}{\begin{eqnarray}}
\newcommand{\eeqa}{\end{eqnarray}}
\newcommand{\beqas}{\begin{eqnarray*}}
\newcommand{\eeqas}{\end{eqnarray*}}
\def\bals#1\eals{\begin{align*}#1\end{align*}}
\def\bal#1\eal{\begin{align}#1\end{align}}

\newcommand{\bcent}{\begin{center}}
\newcommand{\ecent}{\end{center}}

\newcommand{\bitem}{\begin{itemize}}
\newcommand{\eitem}{\end{itemize}}

\DeclareMathOperator{\Tr}{Tr} 

\begin{document}
 
\title{Excess energy and countercurrents after a quantum kick} 

\author{Nuria Santerv\'as-Arranz}  
\altaffiliation[Corresponding author: ]{n.santervas@nanogune.eu}
\affiliation{CIC Nanogune, Tolosa Hiribidea 76, San Sebasti\'an, Spain}
\affiliation{Universidad del País Vasco UPV/EHU, 48080, Bilbao, Spain}

\author{Massimiliano Stengel}
\affiliation{Institut de Ci\`encia de Materials de Barcelona 
(ICMAB-CSIC), Campus UAB, 08193 Bellaterra, Spain}
\affiliation{ICREA-Instituci\'o Catalana de Recerca i Estudis 
Avan\c{c}ats, 08010 Barcelona, Spain}

\author{Emilio Artacho}
\affiliation{CIC Nanogune, Tolosa Hiribidea 76, San Sebasti\'an, Spain}
\affiliation{Donostia International Physics Center DIPC, 
P. Manuel de Lardizabal 4,  San Sebasti\'an, Spain.}
\affiliation{Ikerbasque, Basque Foundation for Science, 48011 Bilbao, Spain}
\affiliation{Theory of Condensed Matter, Cavendish Laboratory, 
University of Cambridge, J J Thomson Avenue, Cambridge CB3 0HE, 
United Kingdom}

\date{\today}

\begin{abstract}
  A quantum system of interacting particles under the effect of a 
static external potential is hereby described as kicked when that 
potential suddenly starts moving with a constant velocity $\mathbf{v}$. 
  If initially in a stationary state, the excess energy at 
any time after the kick equals $\mathbf{v}\cdot\langle \mathbf{P}\rangle(t)$, 
with $\mathbf{P}$ being the total momentum of the system. 
  If the system is finite and remains bound, the long time average 
of the excess energy tends to $Mv^2$, with $M$ the system's total 
mass, or a related expression if there is particle emission. 
  $Mv^2$ is twice what expected from an infinitely smooth onset of
motion, and any monotonic onset is expected to increase the 
average energy to a value within both limits.
  In a macroscopic system, a particle flow emerges countering 
the potential's motion when the particles stay partially behind.
  For charged particles the described kinetic kick is equivalent to
the kick given by the infinitely short electric-field pulse
$\boldsymbol{\mathcal{E}} = \frac{m}{q} \mathbf{v} \, \delta(t)$ 
to the system at rest, useful as a formal limit 
in ultrafast phenomena.
  A linear-response analysis of low-$v$ countercurrents in 
kicked metals shows that the coefficient of the linear term in $v$
is the Drude weight.
  Non-linear in $v$ countercurrents are expected for 
insulators through the electron-hole excitations induced by the 
kick, going as $v^3$ at low $v$ for centrosymmetric ones. 
  First-principles calculations for simple solids are used to 
ratify those predictions, although the findings apply more generally
to systems such as Mott insulators or cold lattices of bosons or
fermions.
\end{abstract}

\maketitle

\section{Introduction}

  Back in 1939, Arkady Beynusovich Migdal \cite{Migdal1939} 
offered a description of the behavior of the electrons 
of an isolated atom when its nucleus suffers an abrupt 
jolt when hit by a neutron.
  Since then, the Migdal effect has appeared in various contexts and
is now quite prominent in the literature around the detection of
dark matter particles (see e.g. Ref.~\cite{darkmatter}).
  It provides predictions for intra-atomic excitation and 
ionization probabilities.
  This work focuses into the effect of such a jolt to the 
total energy and generalizes it to any system of quantum
particles under an external potential which experiences 
such a kick as given by a sudden onset of constant-velocity 
translation motion without deformation.

  Beyond the atomic case, the rigidity of the potential 
is a rather artificial construct in that, if actually hitting 
a molecule or solid, for instance, it will deform under the jolt, 
changing the shape of the potential over time. 
  Our interest was however raised by the fact that for some 
theoretical simulations of physical non-equilibrium processes
such onsets are used.
  That is the case, for instance, for many first-principles 
calculations of electronic stopping processes of ion projectiles
in condensed matter (see e.g. Refs.~\cite{Pruneda2007, Zeb2012, 
Ullah2015, Schleife2015, Ullah2018, Correa2018, Halliday2022}),
and they also appear in the recently proposed theory 
for Born effective charges in metals \cite{Dreyer2022}.
  More generally, however, given the increasing interest in the 
description of ultrafast quantum dynamics phenomena
\cite{Nisoli2017,Hui2024}, the availability of formal results 
should always be of interest.
  Furthermore, besides the single atom case, new experiments 
and/or further theoretical constructs beyond the imagination 
of the authors should be forthcoming for which this work could 
be of direct relevance.
  In particular some of the predictions of this paper might 
be tested in cold-gas \cite{Eigen2018} or cold-lattice 
\cite{Song2022} settings where the kick might be experimentally 
implemented on the potential generated by their lasers.

  Another motivation for our work originates from the intriguing 
parallels that exist between time-dependent (roto)translations and 
electromagnetism.
  The quantum kick explored here can be indeed regarded as the 
mechanical realization of an ultrafast electric field pulse 
applied to a system at rest. 
  As such, the countercurrents observed in the computational
experiments shown below can be understood within the 
well-established framework of linear and nonlinear optics. 
  In particular, we find that their low-velocity limit relates 
to the Drude weight, a defining property~\cite{Resta2018} of 
the metallic state, and possibly the most fundamental parameter 
in the theory of electron transport~\cite{Allen}. 
Our results point to a purely mechanical interpretation thereof, that is,
based on the energetics of the electron system following an instantaneous
boost, and not invoking any coupling to electromagnetic fields.

\section{Formal}

\subsection{Excess energy after quantum kick}

\subsubsection{Excess energy}
   The problem considered is that of a quantum system of
interacting particles under the influence of an external local 
potential
of the form
\beqs 
\mathrm{V} = \sum_j V(\mathbf{r}_j)
\eeqs
$j$ summing over all particles.
  Being the system originally ($t<0$) in a stationary state, 
the quantum kick or jolt is defined by the sudden onset of motion
of the potential at $t=0$. 
  More specifically, at that time the potential starts moving at 
a constant velocity $\mathbf{v}$  
\beq
\label{eq:movepot}
\mathrm{V}  (t\geq 0)= \sum_j V(\mathbf{r}_j-\mathbf{v}t) \, .
\eeq
  This paper is about the excess energy communicated by 
the motion of the potential to the system
\beq
\label{eq:excess}
\Delta E(t) \equiv \langle H\rangle(t) - E_0
\eeq
being $E_0= E(t\!\!<\!\!0)$ the system's energy prior to the kick, 
namely, the time-independent $\langle H\rangle$ for $t<0$.
   The mentioned stationary state can be pure, so that
$\langle H\rangle=\langle \Psi_n | H |\Psi_n\rangle$, being 
$|\Psi_n\rangle$ any eigenstate of the time-independent
Hamiltonian prior to the kick, or a mixed state, where
$\langle H\rangle=\Tr\{\rho H\}$, the density matrix being
$\rho = \sum_n \rho_n |\Psi_n\rangle\langle\Psi_n|$, 
including thermal equilibrium as a particular case.

\subsubsection{Change of reference frame}

   A simple expression for the excess energy is obtained by
using invariance under Galileo transformations.
  As done by Migdal \cite{Migdal1939}, the reference frame 
can be changed from the one seeing the process as described, 
which we will call RF1, to the one of an observer moving with 
a constant velocity $\mathbf{v}$ with respect to the original 
one at all times (RF2).
   For the observer at RF2 the process will be that of the system 
in equilibrium with respect to the potential, which moves with 
$-\mathbf{v}$ at negative times, and suddenly stops at $t=0$, 
the potential remaining immobile at all later times.

  The total energy of the system for RF2 at $t<0$ is 
\beq
\label{eq:eprime}
\langle H \rangle' =  E_0 + \frac{1}{2} M v^2
\eeq
where primed symbols indicate magnitudes corresponding to RF2, 
$M$ is the total mass of all particles in the system, and $v$ 
is the magnitude of the $\mathbf{v}$ vector.
  It is easily obtained by changing from RF1 where the energy was $E_0$.
  It is therefore a constant in spite of the Hamiltonian depending 
explicitly on time in RF2.
  Equation~\eqref{eq:eprime} is easy to generalize allowing for
a stationary state of an extended system with a stationary 
flow of particles (a current), but we will not consider it here
for simplicity.

\subsubsection{Simple relation}
\label{sec:relation}

  For $t>0$ in RF2 the energy remains constant at all times no 
matter how complicated the dynamics may be after the sudden stop, 
since the system dynamics is defined by an initial wave-function 
(or density matrix) at $t=0$ and a time-independent potential at all 
ulterior times.
  The energy therefore remains constant as defined in Eq.~\eqref{eq:eprime}
at all times.
  For RF1 the situation is very different, since the system was 
in a stationary state and its suddenly starting evolution at $t\geq 0$ 
gives a $\langle H\rangle$ that varies in time. 
  It is easy to show that this instantaneous total energy 
at any positive time fulfills the simple relation
\beq
\label{eq:theorem}
\langle H \rangle(t) =  E_0 + \mathbf{v}\cdot \langle \mathbf{P}\rangle (t)
\eeq
where $\mathbf{P}$ is the total momentum of all particles in 
the system (in RF1)
\footnote{In the presence of a stationary current for $t<0$ in an extended
or periodic system, the relation generalizes to $\langle H \rangle(t) =  
E_0 + \mathbf{v}\cdot \Delta\langle \mathbf{P}\rangle (t)$, where
$\Delta\langle \mathbf{P}\rangle (t) \equiv \langle \mathbf{P}\rangle (t) 
- \langle \mathbf{P}\rangle_0$, and where $\langle\mathbf{P}\rangle_0$ 
is the net momentum associated to the mentioned stationary current.}.
  Therefore, as the moving potential starts pushing the particles, they 
acquire a net momentum that raises the energy. 
  And, besides $\mathbf{v}$, \textit{the energy raised depends 
solely on the net momentum acquired.}

  Although it could be shown within RF1 (see Appendix~\ref{app:lagrf1}),
the relation \eqref{eq:theorem} is easily and usefully obtained 
from the transformation of the energy under a Galilean boost 
\cite{Baym2018} to RF2,  
\beq
\label{eq:tot}
\langle \mathrm{H} \rangle ' = \langle \mathrm{H} \rangle - 
\mathbf{v}\cdot\langle \mathbf{P}\rangle  +\frac{1}{2}Mv^2 \, 
\eeq
(although well known, a reminder of the derivation of this 
relation in our context can be found in the Appendix~\ref{app:galileo}).
  Knowing that $\langle \mathrm{H} \rangle '$ is constant at all times 
(including $t>0$), as given by Eq.~\eqref{eq:eprime}, introducing
it in Eq.~\eqref{eq:tot} gives the proposed relation in 
Eq.~\eqref{eq:theorem}, or, for the excess energy in Eq.~\eqref{eq:excess},
\beqs
\Delta E(t) = \mathbf{v}\cdot \langle \mathbf{P}\rangle (t) \, .
\eeqs
  It is simple but it can be quite counter-intuitive.
  A perfectly constant energy in RF2 becomes a time-dependent one in 
RF1, and quite definitely so.

  It is illustrative to take the simplest possible case, a
single harmonic oscillator of mass $m$ and frequency $\omega$ 
in one dimension, initially in its ground state.
  In RF2, at $t=0$ the wavefunction is the ground-state 
Gaussian times $e^{- i m v/\hbar}$, 
which happens to be a coherent state with 
$E'=E_0 +\frac{1}{2}m v^2$. For $t>0$, 
$\langle p\rangle '(t)=-mv \cos\omega t$, giving
$
\Delta E (t\geq 0)= \langle p\rangle v = mv^2 (1-\cos \omega t )
$ 
while it was constant (zero) at any earlier time.
  That marked time-dependence is quite in
contrast with the constant $E_0+mv^2/2$ at all times for RF2.

  $\Delta E$ at positive times averages to $mv^2$, but
peaks at $2mv^2$, twice the average. 
  Richer cases are shown in Section~\ref{sec:numerical}, but
it is worth emphasizing here that, in addition to the contrast
in energy behaviour between RF1 and RF2, in RF1 the kick can give 
an initial push of up to twice the average of $\Delta E$.
  Initial, since, in the presence of more degrees of freedom, 
the excess energy will distribute among various modes, 
dampening the initial oscillation into noise around the 
average.

  A final remark to the time-dependent excess energy: 
the fluctuation in the energy relates to a fluctuating
force exerted on/by the agent keeping the constant velocity
of the potential.
  In RF2 the same force would be acting, but, since 
$\mathbf{v}=0$, there is no energy change,  given that
the power $\partial_t E=\mathbf{F}\cdot\mathbf{v}$ is zero.

\subsection{Long-time average}

\subsubsection{Bound systems}

  Let us consider first the case of a finite system that remains
bound at all times, that is, there is no particle emission.
  In long times after the kick, it is expected that the
average momentum would average to $Mv$.
  Easy to see in RF2, where an initial tendency for 
the particles to continue their motion in the $-\mathbf{v}$
direction should dissipate and thermalize to a long-term 
$\langle \mathbf{P}\rangle'\rightarrow 0$.
  That means that the long-time average of the 
excess energy for a bound system should be
\beq
\label{eq:longtimebound}
\overline{\Delta E} = \frac{1}{\tau}\int_0^{\tau} \! \!
\Delta E(t) \mathrm{d}t \longrightarrow Mv^2 \, .
\eeq
  It is interesting since the same system in the initial 
stationary state displacing with velocity $\mathbf{v}$ 
(as in just changing reference frame, with no kick) would 
have an excess energy of $Mv^2/2$, which is also what would 
be expected from an initial stationary system which 
acquires that velocity in an infinitely smooth fashion.
  It is therefore sensible to conjecture that for a system
that remains bound at all times which is subjected to 
a monotonic increase in velocity from zero to $\mathbf{v}$
should experience an increase in energy between both 
limits, 
\beqs
\frac{1}{2}Mv^2 \, \leq \, \overline{\Delta E} 
\, \leq \, Mv^2 \, ,
\eeqs
one for the infinitely smooth onset, the other for
the infinitely abrupt.

  A simple way of understanding Eq.~\eqref{eq:longtimebound}
is again turning to RF2. 
  There, the system comes from negative times with a
constant excess energy of $Mv^2/2$ and it is suddenly stopped
converting that excess kinetic energy into internal energy.
  Changing back to RF1 gives it an additional excess energy
that averages to an extra $Mv^2/2$.

\subsubsection{Particle emission}

  A kick on a finite system can give rise to 
particle emission, which, in the case of a system 
of electrons (as an atom in the Migdal effect) means
ionization. 
  The excess energy in Eq.~\eqref{eq:theorem} 
is still valid, but it gives the result for the total 
system, including the emitted particles. 
  In this section we are interested in obtaining an 
expression for the excess energy in the system remaining 
after particle emission, which implies redefining the
energy reference.

  Consider first one emitted particle of mass $m$.
  It is assumed to 
fly off freely, with a well defined momentum $\mathbf{p}$
and energy $\varepsilon_p = p^2/2m$.
  Hence, we define the excess energy relevant to the
remaining system as
\beq
\label{eq:def-emission}
\Delta E_s(t) \equiv \Delta E(t) - \varepsilon_p \, .
\eeq
  Similarly, the total momentum decomposes into
the one of the remaining system and the one of
the particle, $\mathbf{P}_s\equiv\mathbf{P} - \mathbf{p}$.
  Using Eq.~\eqref{eq:theorem} and the above definitions
the following relation is easily obtained
\beq
\label{eq:th-emitted}
\Delta E_s(t) = \mathbf{v}\cdot\langle\mathbf{P}_s\rangle(t) 
- \delta \varepsilon
\eeq
where $\delta \varepsilon\equiv \varepsilon_p -
\mathbf{v}\cdot\mathbf{p}$.
  This latter correction can vary between
\beqs 
-\varepsilon_p \, \leq \, \delta \varepsilon \, \leq \, 
3 \varepsilon_p
\eeqs
depending on the direction of emission, the limits
corresponding to the particle being emitted in the direction 
of $\mathbf{v}$ or against it, respectively.

   Following the same arguments as for the bound case,
the long-time average for the remaining system excess energy
would be
\beq
\label{eq:longtime-emission}
\overline{\Delta E}_s \longrightarrow (M-m)v^2 -\delta\varepsilon\, ,
\eeq
since the average momentum in the remaining system 
will thermalise to $\overline{\langle\mathbf{P}_s\rangle}
\longrightarrow (M-m)\mathbf{v}$.

  For a monotonic onset of motion giving rise to particle emission, 
the remaining system would show a long time average excess 
energy
\beq 
\overline{\Delta E}_s \, \leq \, (M-m)v^2 -\delta\varepsilon\, ,
\eeq
where we can now take the more general case of
emission of $N$ particles, defining
\beqs
\delta \epsilon \equiv \sum_i^N \left (\frac{p_i^2}{2m_i} - 
\mathbf{v}\cdot \mathbf{p}_i \right ) \quad
\textrm{and} \quad m\equiv\sum_i^N m_i \, .
\eeqs
$i$ running over the emitted particles.
  Even more generally, the probabilistic quantum-mechanical 
prediction of particle emission would effectively contemplate
a non-integer number of emitted particles, which could also be 
easily incorporated.
  The smooth onset limit remains as $Mv^2/2$ since no 
particles are emitted in the adiabatic limit.

  Note that Eq.~\eqref{eq:def-emission} is a definition
and other energies for the remaining system are possible
(e.g. with explicit consideration of the energy cost of
extracting a particle with zero kinetic energy, the 
ionization potential), which would affect the final form
of the $\delta \epsilon$ correction in Eqs.~\eqref{eq:th-emitted}
and \eqref{eq:longtime-emission}.
  The key point remains that the total excess energy
$\overline{\Delta E}\rightarrow (M-m) v^2 + \mathbf{v}
\cdot \mathbf{p}$, which is in general different from
$Mv^2$.

\subsubsection{Extended systems, countercurrents}

  The situation changes for extended systems, such 
as electrons in infinite or periodic-boundary solids. 
  Metals can easily sustain currents, and it is to be
expected that if starting to move the nuclei rigidly and abruptly
at $t=0$ the electrons could tend to stay behind, thereby 
establishing a current in the solid  in the opposite
direction, a countercurrent.
  It is defined with respect to the solid and, therefore,
in RF2.
  Such a current will be transient and eventually dissipate
away as the current in a metal does after an electric
field has been switched off.
  The current dissipation (in the absence of phonons)
would be due to electron viscosity and drag effects
arising from electron-electron interaction. 
  The time scale for that dissipation, however, is expected
to be significantly larger (picoseconds) than the time 
scale for thermalizing to the defined excess energy 
by other processes (femtoseconds) allowing to characterize 
the countercurrent by suitable averages at intermediate times.

  The excess energy is therefore expected to remain
below $Mv^2$ while the current
is sustained, since $\langle \mathbf{P}\rangle \rightarrow
M \mathbf{v} - m_e \Omega \,\mathbf{j}'/e$ during the time the 
current (of current density $\mathbf{j}'$) lasts, giving
\beq
\label{eq:countercurrent}
\overline{\Delta E} \rightarrow Mv^2 - \frac{m_e \Omega}{e} 
\mathbf{v} \cdot \mathbf{j}'
\eeq
being $m_e$ and $e$ the mass and charge of each electron,
respectively, $\Omega$ being the unit-cell volume, and 
the energy and momentum values being given per unit cell.

  A useful limit for countercurrents is the one given
by a weak periodic potential for the electrons in the
solid, that is, $V_0\gg E_F$, being $V_0$ the amplitude of
that potential and $E_F$ the Fermi energy of the metal.
  In the limiting case, $V_0\rightarrow 0$ the homogeneous 
electron liquid (jellium) is recovered. 
  If the constant external potential of jellium starts
displacing at $t=0$ absolutely nothing happens in the 
system, remaining in the stationary state it had before.
  If $\mathbf{j}=0$ before, it remains unchanged, and
from the point of view of RF2, there is a current density 
\beq
\label{eq:idealcurrent}
\mathbf{j}'_0=- e \, n \, \mathbf{v} = 
-\frac{eN_e}{\Omega} \mathbf{v}
\eeq
describing the ideal countercurrent, with an excess 
energy $\Delta E=0$, as expected (nothing happened),
being $N_e$ the number of electrons per unit cell
when defined, and $\Omega$ its volume.

  For a metal under a non-constant potential, now resorting
to the single-particle language, since the original translational 
invariance of the solid is not affected by the motion, the 
crystal momentum $\mathbf{k}$ remains well defined, but each 
Bloch state has its original crystal momentum shifted to 
$\mathbf{k}' = \mathbf{k}-m \, \mathbf{v}/\hbar$.
  It produces a rigid shift of the Fermi surface, which gives 
the paradigmatic picture of a metal under current.

  At $t=0$ the Bloch states have wavefunctions that 
correspond to the unshifted crystal momentum, and, are 
therefore not eigenstates of the corresponding Hamiltonian,
but rather combinations of states with the same $\mathbf{k}$.
  In other words, they include vertical electron-hole pair
excitations, which then will evolve in time, but still
conserving $\mathbf{k}$ and the Fermi surface shift.

  For insulators, however, although the shift of crystal
momentum remains, the absence of a Fermi surface prevents 
the appearing of a countercurrent as described above (all
Bloch states for filled bands equally shifted in $\mathbf{k}$
has no net effect on crystal momentum).
  A priori one could expect an excess energy 
$\overline{\Delta E}\rightarrow M v^2$ stemming from
vertical electron-hole excitations.
  However, these excitations can provide carriers that 
allow for currents to arise. 
  In Section~\ref{sec:numerical} numerical calculations
are presented for a prototypical metal and insulator.

\subsubsection{Stopping power in jellium}

  As last formal point, we make here a connection with results
for the electronic stopping power $S_e$ of a projectile ion 
shooting through the homogeneous electron liquid. 
  Describing the projectile as an external potential acting on
the jellium electrons that displaces with constant velocity 
$\mathbf{v}$, it establishes a stationary state in which
energy is constantly being transferred by the projectile
motion to the electrons at a rate defined by $S_e$ in
terms of energy per unit of projectile displacement, the
stopping power \cite{Echenique1981}.
  Therefore, the excess energy rises steadily with
$\partial_t \Delta E = S_e v$. 

  The result in Eq.~\eqref{eq:theorem} is applicable and
implies that the projectile is steadily pushing the electrons 
giving the whole system a steadily rising momentum 
$\langle \mathbf{P} \rangle$ in the direction of $\mathbf{v}$ 
such that 
\beq
\label{eq:stopping}
\partial_t P = S_e
\eeq
$P$ being the magnitude of $\langle \mathbf{P}\rangle$.
  It is not a surprising relation if one thinks in terms of 
Ehrenfest theorem and the fact that $S_e$ equals the force 
exerted on the electron system by the moving projectile.
  Notice that in this case the sudden onset has no effect
on the result, since it just represents an initial 
transient with a possible energy shift of finite size 
that does not affect the discussed derivatives for 
the steady state.
  Equation~\eqref{eq:stopping} is directly related to the work for 
$S_e$ in jellium by Echenique \textit{et al.} \cite{Echenique1981},
where they used the same RF1-RF2 transformation to obtain
a theory of stopping power from a theory of resistance 
by impurity scattering in an ideal metal.

\subsection{Relation to electromagnetism}
\label{sec:electromag}

A crystal displacing rigidly with respect to the laboratory frame 
with constant velocity ${\bf v}$ is in all respects equivalent, 
in its own co-moving frame, to a crystal at rest, but with a constant 
vector potential $\mathbf{A}$ acting on the wavefunctions.
In both cases, the orbitals acquire the same complex phase that increases 
linearly along ${\bf A}$ or ${\bf v}$, but all physical properties remain 
the same as if there were neither ${\bf v}$ nor ${\bf A}$. 
This well-known fact is referred to as Galileian invariance and 
electromagnetic (EM) gauge invariance in their respective contexts. 

Rigid boosts or uniform vector potentials have an observable effect 
only when ${\bf A}$ or ${\bf v}$ change in time; and not surprisingly, the 
EM counterpart of an acceleration is an electric field.
Following this analogy, it is easy to show that the quantum kick 
described in this work is the \emph{exact} mechanical realization 
of an electrical pulse in the form $\mathcal{E}(t) =\mathcal{E}_0 
\delta(t)$.
Interestingly, this \emph{Gedankenexperiment} is well known in
the theory of optical response~\cite{Allen,Resta2018}, and referred 
to with essentially the same terminology (``kick'') as we use here.
A kick in terms of an infinitely narrow electrical pulse 
was also proposed in the context of real-time TDDFT that started in 
Ref.~\cite{yabana1996}, which was explicitly formulated as an
impulse initial wavefunction in Ref.~\cite{bertsch2000} (essentially
$\Psi(t=0^+)$ in RF2 in our nomenclature), and which gave a 
very fruitful line of research (see e.g. Ref.~\cite{Schultze2014}
for a prominent example).
It is therefore convenient to interpret our findings 
obtained using $\Delta E$ in RF1 by borrowing
the relevant concepts from the theory of the optical response, which is
in a much more mature state than its mechanical counterpart.

For simplicity we shall initially assume a regime where the kick is small: 
as such, it can be treated perturbatively, and its effects on the 
current discussed within a linear-response framework.
In the frequency domain, the linear optical conductivity $\sigma$ 
relates the current to the electric field via
\begin{equation}
    j(\omega) = \sigma(\omega) \mathcal{E}(\omega).
\end{equation}
In metals, it is most appropriate to separate 
$\sigma=\sigma^{\rm (Drude)} + \sigma^{\rm (reg)}$ into a Drude 
contribution, 
\begin{equation}
    \sigma^{\rm Drude}(\omega) = \frac{i}{\pi} \frac{D}{\omega + i \eta^+},
\end{equation}
where $\eta^+$ is a positive infinitesimal within the undamped regime, 
and a remainder regular part that remains finite in the transport 
($\omega\rightarrow 0$) limit.
$D$ is known as the Drude weight, a defining property of the metallic 
state that accounts for the inertia of the free carriers. 
In an isotropic free-electron gas it corresponds to $D=\pi n/m_e$ 
with $n$ the electron density.
In an insulator, $D=0$ and only $\sigma^{\rm (reg)}$ survives, 
as the latter contains contributions from all the vertical 
interband transitions.

Recall now that a Dirac delta in time has a flat frequency spectrum,
\begin{equation}
 \mathcal{E}(t) =   \frac{\mathcal{E}_0}{2\pi} \int_{-\infty}^{+\infty} \!
 \mathrm{d} \omega \, e^{-i\omega t}.
\end{equation}
Thus, within our present assumptions, at any time the current $j(t)$ 
is given as a Fourier transform of $\sigma$,
\begin{equation}
    j(t) = \frac{\mathcal{E}_0}{2\pi} \int_{-\infty}^{+\infty} \! \mathrm{d}
    \omega \, \sigma(\omega) e^{-i\omega t}.
\end{equation}
The Drude contribution can be integrated analytically, and yields
\begin{equation}
    j^{\rm (Drude)}(t) = \frac{\mathcal{E}_0 D}{\pi} \theta(t),
\end{equation}
where $\theta(t)$ is the Heaviside function: $\theta(t>0)=1$ and 
vanishes for $t<0$.
Remarkably, once the carriers are set in motion by the kick, 
they keep propagating with the same velocity forever, reflecting the 
assumed absence of scattering processes.
Regarding the regular part, we can write
\begin{equation}
\label{jreg}
    j^{\rm (reg)}(t) = \frac{\mathcal{E}_0}{\pi} \theta(t) 
    \int_0^{+\infty} \! \mathrm{d}\omega \, {\rm Re} \, 
    \sigma^{\rm (reg)}(\omega) \cos(\omega t).
\end{equation}
Here the real part of $\sigma^{\rm (reg)}(\omega)$ describes optical 
absorption processes that are due to interband resonances; 
these are typically separated in frequency from the main Drude peak.
Their effect is far from negligible: indeed, immediately after the 
kick the total current is given by the $f$-sum rule,
\begin{equation}
    j(t=0^+) = \frac{\mathcal{E}_0 e^2 n}{m_e},
\end{equation}
where $n=N/\Omega$ is the total electron density in the system 
including low-lying core states. 
$e^2 n/m_e$ is typically much larger than $D/\pi$, as the latter 
only reflects the small fraction of free carriers.
Notwithstanding, the reason why we obtain such a large current 
at $t=0^+$ is that the electrons have just felt the kick, 
but didn't have time to interact with the underlying crystal 
potential yet. 
As they do, all Fourier components of Eq.~\eqref{jreg} start 
oscillating incoherently (the cosines in the integral are in 
phase only at $t=0$) and mediate to zero on average.
Therefore, by looking at the running average of $j(t)$ after 
some time has elapsed, one obtains a reliable measure of 
$j^{\rm (Drude)}$.

The above theory predicts that metals will have a current 
increasing \emph{linearly} with kick velocity $v$, and with 
a constant of proportionality that is given by the Drude weight. 
Conversely, there will be no steady current whatsoever 
in an insulator. 
To obtain a complete picture, however, we must go beyond the 
linear regime and consider the contribution of higher-order 
terms in the electric field. 
Terms of order $E^2$ have to do with the bulk photovoltaic 
effect~\cite{Rappe-review}: in principle, they yield a current 
that goes like $v^2$ at small velocities, and is therefore 
unsensitive to the sign of $v$.
The related nonlinear susceptibility coefficients vanish in 
the presence of space inversion symmetry, though, which means 
that this effect is absent in the materials studied below.
It is easy to see that, with space inversion, the current at 
long times after the kick must go like
\begin{equation}
    j^{\rm (steady)} =   \frac{D}{\pi}\mathcal{E}_0 + A \mathcal{E}^3_0 +  
    B \mathcal{E}^5_0 + \cdots
\end{equation}
i.e., it can only contain even powers of the field. 
The presence or the absence of the linear term, therefore, 
marks the qualitative difference between the behavior of metals 
and insulators; we shall see a convincing demonstration of this 
fact in our numerical tests.
Note that only the linear coefficient has a universal sign: 
it can be either zero or positive; the sign of the higher-order 
coefficients is material dependent. 
Thus, our theory does not exclude ``overshoot'' currents: 
it only does so in the low-energy limit, where the current is 
dominated by the free carriers that --not being bound to the 
lattice-- are ``left behind'' by the kick.

\section{Numerical study of currents}
\label{sec:numerical}

\subsection{Method}
\label{sec:method}

  First-principles calculations were performed
for ideal crystals of aluminum and diamond 
under the effect of a sudden onset of 
constant-velocity motion of the external potential
acting on their valence electrons describing the
interaction with the core ions in the crystal.
  All calculations were performed in RF1.
  Time-dependent density-functional theory \cite{Runge1984} 
was used, propagating the Kohn-Sham wavefunctions in real 
time (rt-TDDFT) by discretizing time and using a 
Crank-Nicolson integrator as implemented \cite{Tsolakidis2002}
in the \textsc{Siesta} program \cite{Soler2002,Garcia2020}.
  An initial ground-state density-functional calculation
was used for each system to establish the static Kohn-Sham
wave-functions as initial ones at $t=0$. 
  The dynamic runs then were performed with all core ions
moving rigidly in the $\langle 100 \rangle$ direction
at various constant velocities.

  Adiabatic PBE \cite{Perdew1996} was used for the 
gradient corrected in space, local in time, exchange-correlation
potential (initial static calculations were performed 
with PBE).
  Adiabatic PBE is not expected to reproduce the correct
dissipation of transient currents, but that would only
affect longer time scales than the ones 
explored in this work.
  The PBE density functional for the energy was used as 
estimator of the evolving expectation value of the total 
energy of the electron system
$\langle H \rangle(t)$ \cite{Halliday2022}.
  
  Hamiltonian matrix elements were calculated
by discretizing real space with the grid fineness corresponding
to a 350 Ry energy cutoff \cite{Soler2002}, and 
reciprocal space was sampled with a grid of \textbf{k} points
corresponding to 18~\AA\ and 19~\AA\ length cutoffs 
\cite{Moreno1992} for diamond and aluminum, respectively.
  The lattice constant was set to 3.57~\AA\ for diamond and 
4.05~\AA\ for aluminum, corresponding to their respective 
experimental values.

  Core electrons were replaced by norm-conserving 
pseudopotentials \cite{Troullier1991} and factorized
as described in \cite{Soler2002}, describing the core ions.
  The core cutoff radii of the pseudopotentials are 1.54 Bohr 
and 2.28 Bohr for all angular momentum channels, for diamond 
and aluminum, respectively.
  It should be noted here that the non-local
pseudopotentials introduced as conventionally done in DFT
to describe the effect of core electrons makes the relation 
in Eq.~\eqref{eq:theorem} not strictly applicable 
since they break Galilean invariance (as also
happens in other contexts \cite{Dreyer2022,Dreyer2018,Stengel2018}).
  It is not dissimilar to what is customarily done with
pseudopotential DFT. 
  The introduction of non-local pseudopotentials 
compromises the applicability of the Hohenberg-Kohn 
theorem itself \cite{hohenbergkohn}, but it is routinely done 
under the argument that DFT is applicable to the original 
problem, and pseudopotentials are used to approximate
the solution of the Kohn-Sham equations once formulated.
  However, the non-locality affects (breaks) several formal
relations that would otherwise be fulfilled in various 
contexts \cite{Dreyer2018}, and Eq.~\eqref{eq:theorem}
represents an additional instance. 
The breaking of energy conservation (in RF2)
had been addressed before in linear response 
\cite{tancogne2020octopus, noda2019salmon, pemmaraju2018velocity},
but also for the non-linear correction of currents
\cite{andrade2018negative}.
  Stengel \textit{et al.} \cite{stengel2025pseudopotentials} 
recently addressed it more generally, recovering Galilean
invariance.
  The correction they propose is taken into account in the 
countercurrents reported below.

  A double-$\zeta$ polarized basis set made of finite-support 
pseudoatomic orbitals was used to expand the evolving Kohn-Sham
states. 
  It was generated as explained in Ref.~\cite{Soler2002} using
an energy shift of 10 meV and a split norm of 0.15. 
  The polarization orbitals (a $3d$ extra shell per atom in
both cases) were generated with the perturbative approach 
\cite{Soler2002}.
  As the basis orbitals were displacing with the atoms, the differential
geometry representation of quantum mechanics in curved manifolds 
\cite{Artacho2017} was used and a gauge potential was added 
to the Hamiltonian in the Crank Nicolson integrator \cite{Halliday2021}.
  A time step of 1 attosecond was used in the discretization,
which gave well converged propagation.

  Atomic-orbital bases face difficulties when
$e^{i\mathbf{p}\mathbf{r}}$ phases are required with corresponding
$\hbar/p$ wavelengths that are shorter than interatomic spacings.
  Tests were performed with a plane-wave basis using the 
Qb@ll program \cite{qball}, with the same fundamental approximations,
except the basis, which included plane waves up to a 100
Ry energy cutoff, and a 4th-order Runge-Kutta propagator
with a time step of 0.2 attoseconds. 
  The pseudopotential for C used with Qb@ll was obtained 
from the Schlipf-Gygi norm-conserving pseudopotential library. 
  It was originally generated using the code by D.R. Hamman 
\cite{Hamann2013}.
  Generation and testing details can be found in Ref. 
\cite{Schlipf2015}.
  Results obtained using atomic-orbital and plane-wave 
methods agree very well at low velocities but start to diverge 
for values beyond 9 \AA/fs (see Appendix~\ref{app:basis}).

\subsection{Results and Discussion}

\begin{figure}[t!] 
\includegraphics[width=0.34\textwidth]{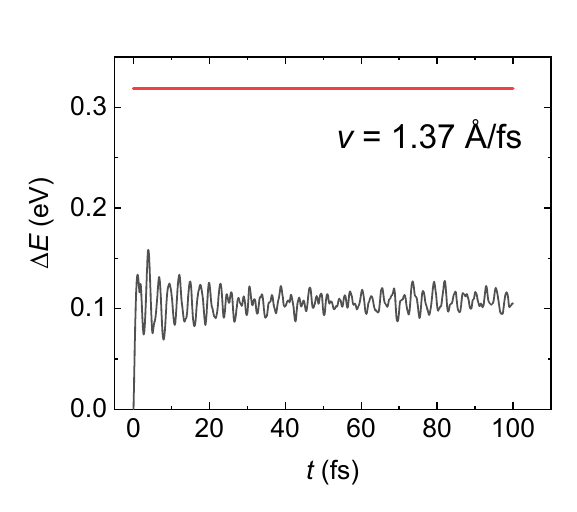} \\
\vspace{-18pt}
\includegraphics[width=0.36\textwidth]{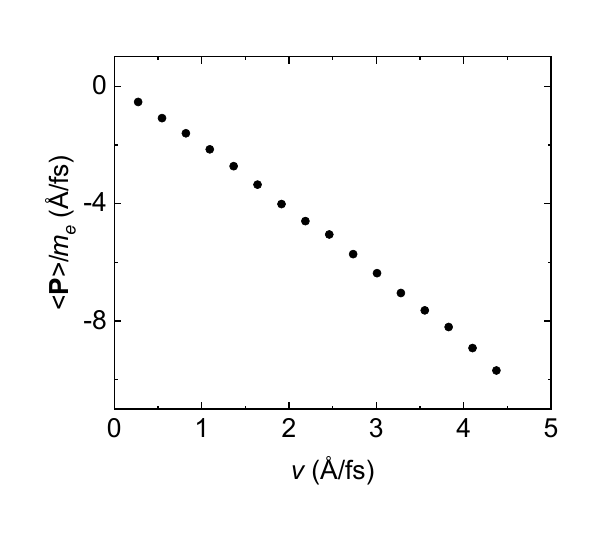} \\
\vspace{-18pt}
\includegraphics[width=0.36\textwidth]{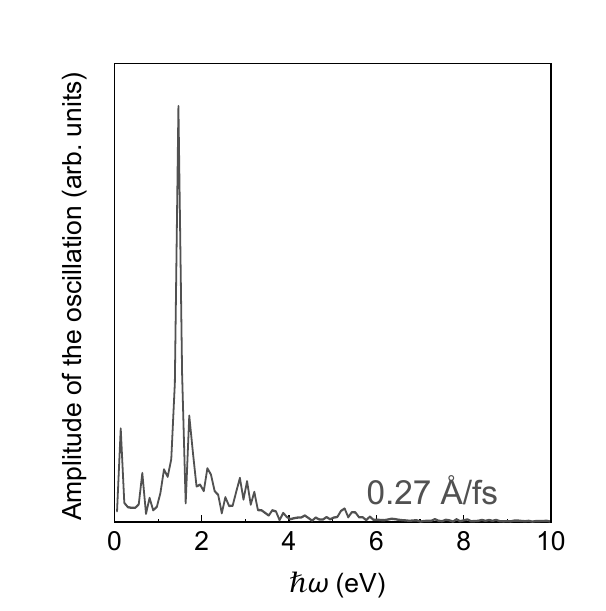} \\
\vspace{-12pt}
\caption{Kicked bulk aluminum. Upper panel: excess energy 
per unit cell $\Delta E$ versus time $t$ after a sudden 
onset of motion with $v=1.37$\AA/fs (black line). 
  The red line gives the theoretical 
reference of $\Delta E=Mv^2$, $M$ being the mass of
the three electrons per cell.
  The difference between the red line and the average 
of the black line indicates the presence of a countercurrent 
[Eqs.~\eqref{eq:countercurrent} and \eqref{eq:counteral}].
  Middle: Magnitude of average momentum (or of current density)
$\langle \mathbf{P}\rangle'/m_e =
\frac{\Omega}{e}\mathbf{j}'$ versus velocity $v$.
  Lower: Fourier transform of $\Delta E(t)$ versus 
frequency $\omega$ for $v= 0.27$ \AA/fs.}
\label{fig:Al-results}
\end{figure}

\begin{figure}[t] 
\includegraphics[width=0.35\textwidth]{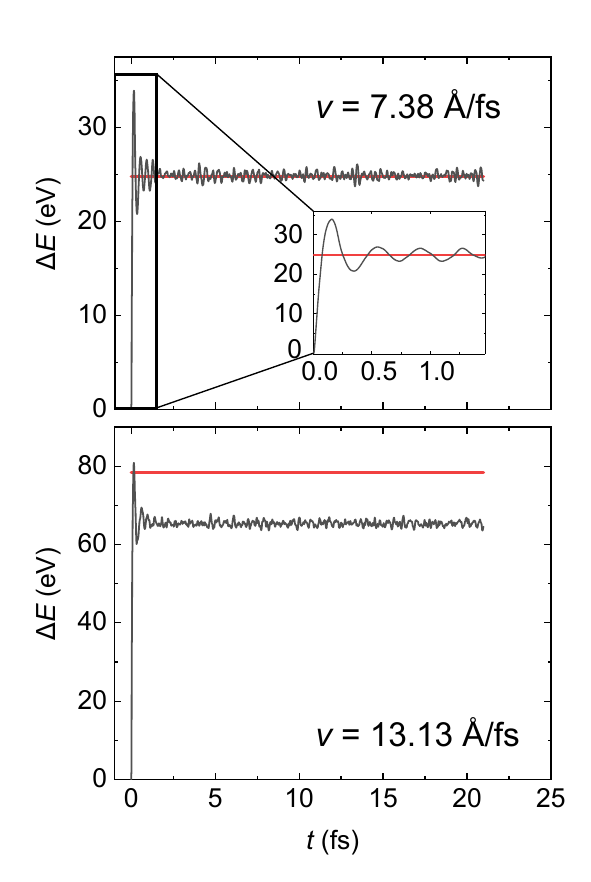} \\
\vspace{-16pt} 
\includegraphics[width=0.35\textwidth]{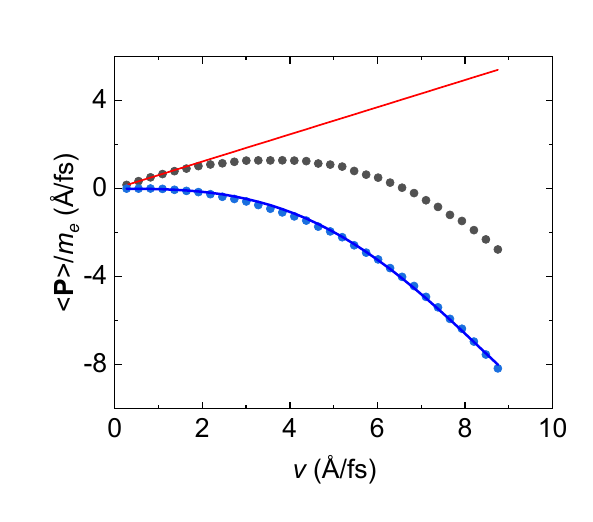}
\caption{Kicked diamond. Upper two panels: excess energy 
per unit cell $\Delta E$ versus time $t$ after a sudden 
onset of motion with $v=7.38$ \AA/fs (inset zoom-in as 
indicated) and $v=13.13$ \AA/fs (black lines). 
  Red lines: theoretical reference 
without countercurrent of $\Delta E=Mv^2$, 
with $M=8 m_e$ for the eight valence electrons per cell.
  Lower: Black dots: magnitude of the average 
momentum per unit cell (or current density) 
$\langle \mathbf{P}\rangle'/m_e =
\frac{\Omega}{e}\mathbf{j}'$ versus velocity $v$ 
from rt-TDDFT results. 
  Red line: linear component at $v\rightarrow 0$, 
due to the non-local pseudopotential 
\cite{stengel2025pseudopotentials}.
  Blue dots: $\langle \mathbf{P}\rangle'/m_e$ after 
removing that linear term.
  Blue line: fit of the latter as
$\langle \mathbf{P}\rangle'/m_e = a v^3 + b v^5$.}
\label{fig:Dia-results}
\end{figure}

  Figures \ref{fig:Al-results} and \ref{fig:Dia-results}
show the results for kicks on perfect crystals of 
aluminum and diamond, respectively, at different 
velocities, where the potential due to the presence of
the ion cores is rigidly displaced at constant velocity
after $t>0$.
  The upper panel of Fig.~\ref{fig:Al-results} shows
the time evolution of the excess energy per unit cell for 
Al for $v=1.37$ \AA/fs (black line).
  The red line indicates the nominal average value of 
$Mv^2$ in the absence of countercurrent, $M$ being 
the mass of the three valence electrons, $3m_e$. 
  The average is clearly below the $Mv^2$ reference
indicating the expected partial countercurrent.
  The magnitude of the countercurrent density can be 
quantified as
\beq
\label{eq:counteral}
j' = \frac{e}{m_e v \, \Omega}(\overline{\Delta E} - Mv^2)
\eeq
using Eq.~\eqref{eq:theorem} and the definition of
$\overline{\Delta E}$ in Eq.~\eqref{eq:longtimebound},
$\Omega$ being the unit cell volume.
  Fig.~\ref{fig:Al-results} (middle) shows it (as 
$\langle p \rangle/m_e$ for velocity units) as a function
of kick velocity.
  A quite linear behavior is observed with a negative 
slope of $2.2$, meaning the countercurrent is 74\% of
the one for the ideal metal limit, $j_0'$.
  It fits quite well with the $\sim 2.0$ carriers
obtained from the charge stiffness (Drude weight) calculated 
in a different way but with similar DFT techniques 
\cite{Dreyer2022}.

  Fig.~\ref{fig:Al-results}(lower) shows the Fourier transform
of $p(t)$ for $v=0.27$ \AA/fs. 
  The well defined oscillation in the upper panel,
especially at the earliest times, is related  
to the main peak in the Fourier transform, which corresponds 
to the electron-hole inter-band excitations in the Al bands
that disperse quite parallel to each other at a vertical 
distance of around 1.5 eV \cite{Ehrenreich1963}.
  It corresponds to the main peak in the optical 
conductivity observed for bulk Al \cite{Ehrenreich1963}.
  Indeed, the Fourier transform of the countercurrent 
offers a way of obtaining $\sigma(\omega)$, both the linear
and non-linear conductivity, analogously to other 
numerical techniques used for the purpose (see e.g.
Ref.~\cite{yabana1996,Tsolakidis2002,Schultze2014}).
  A quantitative use, however, cannot be trusted before 
addressing the problem with non-local pseudopotentials 
mentioned in Section~\ref{sec:method}.

  The two upper panels of Fig.~\ref{fig:Dia-results} 
show $\Delta E$ versus time $t$ for 
diamond for two different velocities.
  The general behavior is similar to what discussed for 
aluminum, except for the fact that the deviation from
the reference excess energy of $Mv^2$ is small
at the lower velocitiy.
  The higher velocity shows a countercurrent, not unexpected
considering the effective doping produced by the very
sizable excess energy provided by the kick, 
and consistent with the discussion in 
Section~\ref{sec:electromag}.

  When plotting the current density as a function of 
kick velocity (lower panel of Fig.~\ref{fig:Dia-results})
a less expected behavior is observed.
  Although on a smaller scale than the currents
in Al, at low kick velocities an overshoot (positive)
current is observed.
  That counterintuitive maximum is not consistent with
what obtained Section~\ref{sec:electromag}, where the linear
term at low $v$ associates to the Drude weight, which should
be non-negative (non-positive countercurrents in our setting),
and, in particular, zero for insulators.  
  It is due to the fact that Eq.~\eqref{eq:theorem} is 
not strictly fulfilled in the presence of the non-local 
pseudopotentials.
  When removing the linear component at the origin from 
the data as an ad hoc correction of that deviation, 
matching the correction proposed in 
Ref.~\cite{stengel2025pseudopotentials},
the remaining curve fits very nicely with 
$j \sim a v^3 + b v^5$ as expected for a centrosymmetric insulator.
(Section~\ref{sec:electromag}).
  Overshoot currents are not impossible and could be understood 
in terms of carriers being generated by the kick but after 
the initial push, given the significant excess energy above the 
reference seen in the upper panel of Fig.~\ref{fig:Dia-results}.
  But no linear term with $v$ would be expected except for the one
associated to the Drude weight (only for metals and of opposite sign).

\section{Conclusions}

  A simple relation between excess energy and momentum
after a quantum system has suffered a sudden onset 
of constant-velocity motion allows obtaining 
interesting predictions about the long-term behavior 
of such kicked systems.
  Formal results include an expected long-time excess
energy average for finite systems, which is conjectured to 
represent an upper bound for any monotonic onset of motion, 
with simple expressions for both with and without particle 
emission.
  In short times after the onset the excess energy can 
rise to up to twice the long-term average, an upper bound 
that would be reached if the single collective mode 
initially excited were not to start thermalizing among
other excitations of the system.

  The connection with momentum has allowed us to
address the possible currents (countercurrents) arising
from kicks in extended systems by monitoring their 
total energy.
  Such currents are studied computationally by 
first-principles calculations of the energy in bulk 
aluminum and diamond, as canonical metal and 
insulator, respectively, by following the dynamics 
of their electrons after a kick using real-time TDDFT.

  The expected countercurrents in the metal are obtained.
  When quantified, they show a current density that is linear 
with velocity, representing a sizable fraction of that 
for the ideal countercurrent and which corresponds
to a good approximation of what expected from the 
Drude weight of $\sim 2.0$ carriers per unit cell \cite{Dreyer2022}.
  The insulator shows the expected non-linear 
countercurrents due to the vertical electron-hole excitations
produced by the kick, which go as $v^3$ for centrosymmetric
diamond (once the spurious linear term is removed).

  Although illustrated with calculations of a simple metal
and a simple insulator, the formal results of this 
work should apply quite generally, and might be of
interest in lattices of cold atoms, both bosons and fermions,
including Hubbard (or Bose-Hubbard) insulators.

\begin{figure}[t] 
\includegraphics[width=0.42\textwidth]{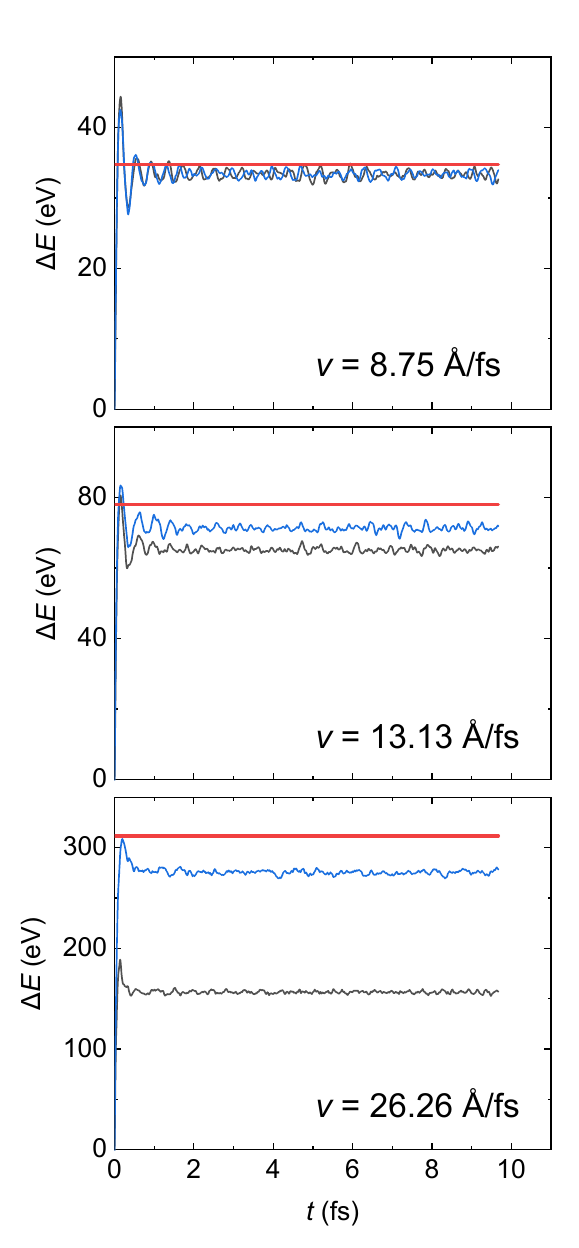}
\caption{Comparison of atomic-orbital and plane-wave 
results for the evolution of $\Delta E$ for diamond
as kicked with the three velocities indicated. 
  The red straight line is the $Mv^2$ reference,
and the black (blue) line is as obtained with an
atomic-orbital (plane-wave) basis.}
\label{fig:basis}
\end{figure}

\begin{acknowledgments}
  We thank F\'elix Fern\'andez-Alonso for bringing to our attention
the literature on the Migdal effect and its application to dark
matter detection, Natalia Koval for technical help setting
up the calculations, and Daniel S\'anchez-Portal,
Rafi Ullah and Daniel Hernang\'omez for 
useful discussions.
  This project is partially supported by the European 
Commission Horizon MSCA-SE Project MAMBA (Grant No. 101131245). 
  Funding from the Spanish MCIN/AEI/10.13039/501100011033 
through grants PID2019-107338RB-C61, PID2022-139776NB-C65, and
PID2023-152710NB-I00,
and FPI grant PRE2022-101273, as well as a Mar\'{\i}a de 
Maeztu award to Nanogune, Grant CEX2020-001038-M 
and a Severo Ochoa Excelence award to ICMAB, Grant
CEX2023-001263-S are also acknowledged, as well as the
Grant No. 2021 SGR 01519 of the  Generalitat de Catalunya,
and the United Kingdom's EPSRC Grant no. EP/V062654/1.
  The authors acknowledge the technical and human support 
provided by the DIPC Supercomputing Center.
\end{acknowledgments}

\appendix  

\section{Excess energy relation from RF1}
\label{app:lagrf1}

  We proved Eq~\eqref{eq:theorem} in Section~\ref{sec:relation} 
using a Galilean boost to RF2 because it is easy, it allows for
an intuitive interpretation of Eq.~\eqref{eq:longtimebound}, and it gives a
direct connection to the physics of ultrashort electromagnetic pulses
as presented in Section~\ref{sec:electromag}.
  It could be done within RF1 as well, however.
  Here we outline a proof in the classical limit, as kindly provided
by a referee.

  Considering the potential in Eq.~\eqref{eq:movepot}, from Newton's 
law, $m_j \ddot{\mathbf{r}}_j = -\nabla_j V(\mathbf{r}_j -\mathbf{v}t)$, 
we get
\bals 
m_j \dot{\mathbf{r}}_j & \cdot \ddot{\mathbf{r}}_j =
- \dot{\mathbf{r}}_j \cdot \nabla_j V(\mathbf{r}_j -\mathbf{v}t) \\
&=- (\dot{\mathbf{r}}_j - \mathbf{v}) \cdot \nabla_j V(\mathbf{r}_j -\mathbf{v}t)
- \mathbf{v} \cdot \nabla_j V(\mathbf{r}_j -\mathbf{v}t) \, .
\eals
  Summing over all particles and integrating over time from $t=0$, it becomes
\beqs
T(t)-T(0) = -(V(t)-V(0)) + \mathbf{v}\cdot \mathbf{P}(t) \, ,
\eeqs
which gives Eq.~\eqref{eq:theorem}, having used $\mathbf{P}(0)=0$.

\section{Transformed energy for mixed state}
\label{app:galileo}

  Considering pure states, to arrive to the Galileo transformation 
of the energy in Eq.~\eqref{eq:eprime} we start by noting that
\beq
\label{eq:pot}
\langle \mathrm{V} \rangle ' \equiv \langle \Psi ' | \mathrm{V}' 
| \Psi ' \rangle = \langle \mathrm{V} \rangle \equiv \langle \Psi  
| \mathrm{V} | \Psi  \rangle \, ,
\eeq
that is to say that the average potential energy (at any time) 
is the same in both reference systems.
  The same Galileo boost transforms the kinetic energy as
\beq
\label{eq:kin}
\langle \mathrm{T} \rangle ' = \langle \mathrm{T} \rangle - 
\mathbf{v}\cdot\langle \mathbf{P}\rangle  +\frac{1}{2}Mv^2 \, .
\eeq
  It is best seen in the real-space representation of $|\Psi\rangle$,
namely, $\Psi(\{s_j\})$, where $s_j\equiv(\mathbf{r}_j \sigma_j)$, 
being $\mathbf{r}_j$ and $\sigma_j$ the position and spin of 
particle $j$, respectively.
  The operators do not change, $\mathrm{T}=\mathrm{T}'$, and the 
boosted wave-function transforms as 
\beqs
\Psi' = e^{i\Omega/\hbar} \Psi \, , \quad \textrm{ where } \quad
\Omega \equiv \sum_j m_j \mathbf{v}\cdot \mathbf{r}_j \, .
\eeqs
  Applying the kinetic energy operator on $\Psi'$
leads directly to Eq.~\eqref{eq:kin}.

  That was valid for any pure state, including time-dependent.
  The generalization to a mixed state is straightforward
using now $\langle \hat{\textrm{O}}\rangle = \Tr\{\rho \hat{\textrm{O}}\}$
for any operator $\hat{\textrm{O}}$, with $\rho = \sum_n \rho_n |\Psi_n\rangle
\langle\Psi_n|$, and with both the basis $\{|\Psi_n\rangle\}$ and the $\rho_n$
coefficients possibly time-dependent.
  Using $\langle \hat{\textrm{O}}\rangle = \sum_n \rho_n 
\langle \Psi_n |\hat{\textrm{O}}|\Psi_n\rangle$, the relations
in Eqs.~\eqref{eq:pot} and \eqref{eq:kin}, can be seen to remain valid
for such mixed states.
  In this paper we use time-independent states (pure and mixed)
for $t<0$, since the more general time-dependent behavior 
leads to a less useful relation.

\section{Basis set dependence}
\label{app:basis}

  Figure~\ref{fig:basis} shows a comparison of the evolution
of the excess energy $\Delta E(t)$ for diamond as obtained 
using \textsc{Siesta} and an atomic-orbital basis versus 
what obtained with Qb@ll using plane waves.
  For $v=8.75$ \AA/fs (upper panel) the agreement is perfectly
sufficient for the purposes of this study.
  It clearly deteriorates for the higher velocities shown in 
the figure.
  The results presented in this work are for velocities 
below 9 \AA/fs, which corresponds to a wavelength 
$\lambda = h / (m_e v)$ of 8.1 \AA, which are
easy to describe with atomic orbitals, with $\lambda/2$
corresponding to interatomic distances in the direction
of motion.
  Higher velocities demand phases varying within atomic sizes,
which would require traveling-orbital bases (see e.g. 
Ref.~\cite{Kimura1987}) or a much larger atomic basis with 
higher angular momenta.
  For the lower velocities used here, the phases vary at 
longer distances, well describable with atomic orbitals 
and complex coefficients.

\end{document}